\documentclass{IEEEtran}
\usepackage{cite}
\usepackage{amsmath,amssymb,amsfonts}
\usepackage{algorithmic}
\usepackage{graphicx}
\usepackage{textcomp}
\def\BibTeX{{\rm B\kern-.05em{\sc i\kern-.025em b}\kern-.08em
    T\kern-.1667em\lower.7ex\hbox{E}\kern-.125emX}}
    \newtheorem{Thm}{Theorem}
\newtheorem{Asu}{Assumption}
\newtheorem{Lem}{Lemma}
\newtheorem{Rem}{Remark}
\newcommand{\g}{{\rm g}}
\newcommand{\diag}{{\rm diag}}
\newcommand{\rank}{{\rm rank}}
\usepackage{graphicx}
\usepackage{color}
\begin{document}
\title{Distributed Consensus of Nonlinear Multi-Agent Systems With Mismatched Uncertainties and Unknown High-Frequency Gains \\(Extended Version)}
\author{Gang Wang, Chaoli Wang, Zhengtao Ding, and Yunfeng Ji
% \thanks{This paragraph of the first footnote will contain the date on 
% which you submitted your brief for review. It will also contain support 
% information, including sponsor and financial support acknowledgment. For 
% example, ``This work was supported in part by the U.S. Department of 
% Commerce under Grant BS123456.'' }
% \thanks{The next few paragraphs should contain 
% the authors' current affiliations, including current address and e-mail. For 
% example, F. A. Author is with the National Institute of Standards and 
% Technology, Boulder, CO 80305 USA (e-mail: author@boulder.nist.gov). }
% \thanks{S. B. Author, Jr., was with Rice University, Houston, TX 77005 USA. He is 
% now with the Department of Physics, Colorado State University, Fort Collins, 
% CO 80523 USA (e-mail: author@lamar.colostate.edu).}
\thanks{G. Wang and Y. Ji are with the Institute of Machine Intelligence, University of Shanghai for Science and Technology, Shanghai 200093,
China. G. Wang was with the Department of Electrical and Biomedical Engineering, University of Nevada, Reno, NV 89557, USA (e-mail: gwang@usst.edu.cn).}
\thanks{C. Wang is with the Department of Control Science and Engineering, University of Shanghai for Science and Technology, Shanghai 200093, China (e-mail: clwang@usst.edu.cn).}
\thanks{Z. Ding is with the School of Electrical and Electronic Engineering,
University of Manchester, Sackville Street Building, Manchester M13 9PL,
UK (e-mail: zhengtao.ding@manchester.ac.uk).}
}

\maketitle

\begin{abstract}
This brief addresses the distributed consensus problem of nonlinear multi-agent systems under a general directed communication topology. Each agent is governed by higher-order dynamics with mismatched uncertainties, multiple completely unknown high-frequency gains, and external disturbances. The main contribution of this brief is to present a new distributed consensus algorithm, enabling the control input of each agent to require minimal information from its neighboring agents, that is, only their output information. To this end, a dynamic system is explicitly constructed for each agent to generate a reference output. Theoretical and simulation verifications of the proposed algorithm are rigorously studied to ensure that asymptotic consensus can be achieved and that all closed-loop signals remain bounded.
\end{abstract}


\section{Introduction}
Distributed control of multi-agent systems has attracted considerable attention over the last two decades, triggered mainly by its wide potential applications and theoretical challenges. As a fundamental problem within the scope of distributed control, designing consensus algorithms, according to which all agents strive to reach an agreement on certain states of interest using only local interactions, has become an active research branch \cite{renbook2008distributed,fu2017finite,wen2014distributed}. In many applications (e.g., course control of marine vessels \cite{fossen2011handbook} and combustion control systems \cite{illingworth2008adaptive}), the high-frequency gain and even its sign may not be known \textit{a priori}. 

For individual systems with unknown high-frequency gains, the Nussbaum gain technique presented in \cite{nussbaum1983some} has been successfully applied to deal with the control problem \cite{xudong1998adaptive,ge2004adaptive}. In case the boundaries of the gains are available, adaptive consensus algorithms have been proposed in \cite{chen2014adaptive} for first- and second-order agents with unknown high-frequency gains. In \cite{ding2015adaptive}, the consensus output regulation problem has been addressed without any knowledge of high-frequency gains. 
In order to deal with uncertain mismatched nonlinear multi-agent systems in the strict-feedback form, distributed adaptive backstepping control strategies have been presented in \cite{peng2014distributed,wangwer2015decentralized,Huang2020}. Within the framework of the prescribed performance technique, some research methods have emerged to solve the distributed leader-following control problem for nonlinear multi-agent systems with mismatched uncertainties \cite{wanggang2017fully,wei2016}. For higher-order nonlinear multi-agent systems with unknown control gains, adaptive consensus approaches have been developed by introducing novel Nussbaum functions in \cite{chen2017adaptive,pengj2018cooperative}. Nevertheless, these algorithms \cite{chen2014adaptive,ding2015adaptive,chen2017adaptive,pengj2018cooperative} require that the signs of high-frequency gains are unknown but identical. Recently, such a requirement has been removed in \cite{wangqing2018adaptive,wang2019distributed}, where a sub-Lyapunov function candidate was skillfully constructed for each agent to analyze the stability of the closed-loop system. Note that these aforementioned results \cite{wangqing2018adaptive,wang2019distributed} can be applied only to a relatively simple class of nonlinear multi-agent systems, where each agent satisfies the matching condition and contains only one unknown high-frequency gain. However, various practical systems do not meet the condition and may involve multiple unknown high-frequency gains, as shown in \cite{xudong1998adaptive,ge2004adaptive}. Another restrictive assumption typically made on higher-order nonlinear multi-agent environments is that all states of the neighboring agents must be available for use in the control law implementation of each agent. This assumption presents a formidable challenge when only the outputs of neighbors can be measured. 

This brief investigates the consensus problem of nonlinear multi-agent systems with mismatched uncertainties and multiple unknown high-frequency gains under general directed graphs. 
% A new dynamic reference output for each agent is presented by invoking the outputs of neighbors. In virtue of this reference output, a distributed consensus algorithm is proposed, such that the control input of each agent does not acquire any additional information from its neighboring agents in addition to their output states. 
The main differences between our work and the existing results can be emphasized as follows.
(i) Compared with previous works on consensus with unknown high-frequency gains \cite{chen2014adaptive,pengj2018cooperative,chen2017adaptive,wangqing2018adaptive,wang2019distributed}, we consider a more general multi-agent system in which each agent is described by higher-order nonlinear dynamics with the mismatched condition, multiple unknown high-frequency gains, and unknown external disturbances.
(ii) Contrary to \cite{chen2017adaptive,chen2014adaptive,pengj2018cooperative,wang2019distributed,wangqing2018adaptive}, the control input can be derived for each agent without requiring any additional information from its neighboring agents other than their outputs, thereby significantly alleviating the communication load in a multi-agent system. Furthermore, the communication graph in this brief is only assumed to have a directed spanning tree. This assumption is less stringent than the undirected connected graph \cite{chen2014adaptive,pengj2018cooperative} and the strongly connected graph \cite{chen2017adaptive,wangqing2018adaptive}. (iii) It should be noted that the solutions in \cite{pengj2018cooperative,wangwer2015decentralized} require each agent to know some prior information on the dynamics of its neighbors such that the adaptive updating laws can be explicitly designed to estimate the unknown dynamics parameters related to the neighbors. In contrast, no preliminary knowledge of the neighbors' dynamics is needed in our work. Moreover, it is no longer necessary for the agent to account for the uncertainties associated with the dynamics of the neighbors.

% From the Lyapunov stability theorem and matrix theory, it is shown that the stability of the closed-loop system and asymptotic consensus among all agents can be guaranteed. 

\emph{Notation:} Throughout the brief, we denote with $1_m$ and $0_m$, respectively, the $m-$vector of all ones and all zeros, and we let $I_m$ denote the $m-$dimensional identity matrix. For a vector function $u(t)$, it is said that $u\in\mathcal{L}_{\infty}[0,t_f)$, if $\sup_{0\leq t<t_f}\|u(t)\|<\infty$ and $u\in\mathcal{L}_{p}[0,t_f)$, if $(\int_0^{t_f}\|u(t)\|^pdt)^{1/p}<\infty$, $p=1,2$. Let $y^{(n)}$ denote the $n$th derivative of $y$.

% We use $\diag\{k_{i}\}$ to represent the diagonal matrix with diagonal entries $k_{1}$ to $k_n$. 

\section{Preliminaries and problem statement}
\subsection{Graph Theory}
In this work, a weighted directed graph $\mathcal{G}=(\mathcal{V},\mathcal{E})$ with the node set $\mathcal{V}=\{1,\dots,n\}$ and the edge set $\mathcal{E}\subseteq \mathcal{V}\times \mathcal{V}$ is used to describe the communication topology among the $n$ agents. An edge $(i,j)\in\mathcal{E}$ indicates that node $j$ has access to the information of node $i$, and node $i$ is a neighbor of node $j$. The set of all neighbors of node $i$ is denoted by $\mathcal{N}_i$. A directed path from node $i_1$ to node $i_p$ is a sequence of ordered edges in the form of $(i_m,i_{m+1}),$ $m=1,\dots,p-1$. A directed graph is said to contain a directed spanning tree if there exists at least a node such that the node has directed paths to all other nodes in $\mathcal{G}$. The weighted adjacency matrix ${A}_n=[a_{ij}]\in R^{n\times n}$ associated with $\mathcal{G}$ is defined by $a_{ij}>0$ if $(j,i)\in \mathcal{E}$, and $a_{ij}=0$ otherwise. The Laplacian matrix ${L}_n\in R^{n\times n}$ associated with $\mathcal{G}$ is defined as ${L}_n={D}_n-{A}_n$, where ${D}_n=\diag\{d_1,\dots,d_n\}$ is the in-degree matrix with $d_i=\sum_{j=1}^na_{ij}$ being the weighted in-degree of node $i$.

% \begin{Lem}\label{Lem_graph}\cite{renbook2008distributed}
% Assume that a matrix $\mathcal{L}_n=[l_{ij}]\in R^{n\times n}$ takes the form of a Laplacian matrix, i.e., $l_{ij}\leq 0$, $i\neq j$ and $\sum_{j=1}^nl_{ij}=0$, $i=1,\dots,n$. Then, the following three conditions are equivalent: (i) the rank of $\mathcal{L}_n$ is $n-1$; (ii) the directed graph of $\mathcal{L}_n$ contains a directed spanning tree; and (iii) $\mathcal{L}_n$ has a single zero eigenvalue, and all other eigenvalues have positive real parts.
% \end{Lem}

\subsection{Problem Statement}
Consider a multi-agent system with $n$ agents, labeled as agents 1 to $n$, under a directed interaction topology. The dynamics of the $i$th, $i=1,\dots,n$, agent is described by
\begin{equation}\label{agent}
\begin{array}{cll}
\dot{x}_{i,\ell}&=&\g_{i,\ell}x_{i,\ell+1}+\theta_{i}^T\varphi_{i,\ell}(\bar{x}_{i,\ell})+\tau_{i,\ell}(t) \\
\dot{x}_{i,m}&=&\g_{i,m}u_i+\theta_{i}^T\varphi_{i,m}(\bar{x}_{i,m})+\tau_{i,m}(t)
\end{array}
\end{equation}
where $\ell =1,\dots,m-1$, $\bar{x}_{i,p}=[x_{i,1},\dots,x_{i,p}]^T\in R^p$ for $p=1,\dots,m$, $u_i\in R$ and $x_{i,1}\in R$ are, respectively, the control input and output of the $i$th agent. $\g_{i,p}\in R$ are the high-frequency gains of the agent, $\tau_{i,p}(t)$ denote uncertain time-varying disturbances evolving in $R$, $\theta_{i}\in R^{p_i}$ is the constant vector of uncertain system parameters, and $\varphi_{i,p}(\bar{x}_{i,p}): R^p\rightarrow R^{p_i}$ are known smooth nonlinear function vectors.

\begin{Rem}
% The agent dynamic equation (\ref{agent}) can represent many models of nonlinear systems; see, for example, single-link manipulators \cite{dawson1994integrator}, vessel maneuvering systems \cite{fossen2011handbook,du2014adaptive}, and jet engines \cite{krstic1995nonlinear}. Therefore, the agent (\ref{agent}) under the directed graph can describe a multi-agent system composed of diverse practical applications. 
It is pointed out that the considered multi-agent system model (\ref{agent}) is more general than the model in most of the currently available results on distributed consensus with unknown high-frequency gains \cite{chen2014adaptive,pengj2018cooperative,chen2017adaptive,wangqing2018adaptive,wang2019distributed}  in the following respects: (i) mismatched uncertainties, multiple unknown high-frequency gains, and uncertain disturbances exist simultaneously in agent dynamics; (ii) the signs of high-frequency gains are allowed to be completely unknown and non-identical.
\end{Rem}

Our control objective is to design a new consensus algorithm for agents (\ref{agent}), based only on their states and the output information of their neighbors, such that (i) all agents can reach asymptotic consensus on the output state, i.e., $\lim_{t\rightarrow\infty}(x_{i,1}(t)-x_{j,1}(t))=0$ for all $1\leq i\neq j\leq n$, and (ii) all signals in the closed-loop system are bounded.

\begin{Asu}\label{asu0}
The high-frequency gains $\g_{i,\ell}$, $i=1,\dots,n,$ $\ell=1,\dots,m$ are unknown and nonzero constants. Besides, there exist unknown positive constants $\tau_{i,\ell}^*$ such that the inequalities $|\tau_{i,\ell}(t)|\leq \tau_{i,\ell}^*$ hold for all $t\geq 0$. 
\end{Asu}

Assumption \ref{asu0} is quite common in the consensus literature \cite{xudong1998adaptive,chen2017adaptive}. The following result can be obtained directly by applying elementary row operations.
\begin{Lem}\label{Lemma1}
Consider a block matrix $E=\left[\begin{array}{ll}
A&B\\
C&D
\end{array}\right]\in R^{m\times n}$. If the matrix $A\in R^{\ell\times\ell}$ with $1\leq \ell\leq \min\{m,n\}$ is nonsingular, then $\rank(E)=\rank(A)+\rank(D-CA^{-1}B)$.
\end{Lem}

To deal with the unknown high-frequency gains of the agents, Nussbaum gain functions $N(\cdot)$ that have the properties that $\lim_{k\rightarrow\infty}\inf\frac{1}{k}\int_0^kN(s)ds=-\infty$ and $\lim_{k\rightarrow\infty}\sup\frac{1}{k}\int_0^kN(s)ds=\infty$ are applied \cite{nussbaum1983some}. In this brief, we utilize the Nussbaum gain function as $N(k)=k^2\cos(k)$.
% \begin{Lem}\label{stability}
% Let $V(t)\geq 0$ and $k(t)$ be smooth functions defined on $[0,t_f)$, and $b$ be a nonzero constant. If the inequality $V(t)\leq c+\int_0^t(bN(k(\sigma))+\mu)\dot{k}(\sigma)d\sigma$ holds for all $t\in[0,t_f)$, where $c$ and $\mu$ are positive constants, then we can conclude that $V(t),$ $\int_0^t(bN(k(\sigma))+\mu)\dot{k}(\sigma)d\sigma,$ and $k(t)$ are bounded over $[0,t_f)$. 
% \end{Lem}

\section{Consensus algorithm design and main result}
% In this section, we present an adaptive control method to realize consensus of the $n$ agents with unknown high-frequency gains. 
Prior to beginning development, we separate the node set $\mathcal{V}$ into two subsets as $\mathcal{V}_1$ and $\mathcal{V}_2$, where $\mathcal{V}_1=\{i\in\mathcal{V}|d_i\neq 0\}$ and $\mathcal{V}_2=\{i\in\mathcal{V}|d_i=0\}$ with $d_i$ being the weighted in-degree of the $i$th agent. Considering that only the output information of neighbors is available for each agent, we propose the following novel dynamic system to generate a reference output $\xi_{i,1}$ for the $i$th ($i\in\mathcal{V}_1$) agent as
\begin{equation}\label{observer}
\begin{array}{lll}
\dot{\xi}_{i,\ell}=\xi_{i,\ell+1}, \dot{\xi}_{i,m}=\gamma_{i}\sum_{j=1}^na_{ij}x_{j,1}-\sum_{p=1}^m\lambda_{i,p}\xi_{i,p}
\end{array}
\end{equation}
where $\ell =1,\dots,m-1,$ the design parameters $\lambda_{i,1},\dots,\lambda_{i,m}$ are positive constants and are selected such that the roots of the characteristic equation $s^m+\lambda_{i,m}s^{m-1}+\cdots+\lambda_{i,1}=0$ are negative real numbers, and $\gamma_i$ is chosen as $\gamma_i=\lambda_{i,1}/d_i$. Since the agents in $\mathcal{V}_2$ cannot receive any information from other agents, the reference output $\xi_{i,1}$ for the $i$th ($i\in\mathcal{V}_2$) agent is developed as $\xi_{i,1}=\gamma_i$ with $\dot{\xi}_{i,1}^{(\ell)}=\xi_{i,\ell+1}=0$ for $\ell=1,\dots,m$, where $\gamma_i$ is a constant chosen arbitrarily by the designer.

% \begin{Rem}
% The strategy behind higher-order dynamics (\ref{observer}) is to produce a continuously differentiable reference output using the output states from neighboring agents. In \cite{rezaee2015average} and \cite{abdessameud2018distributed}, the consensus problems of high-order multi-agent systems have been considered by employing agents' relative position states. Nevertheless, these well-known consensus control methods are limited to linear multi-agent environments. In addition, unknown high-frequency gains and mismatched uncertainties in agents' dynamics have not been considered in controller design.
% \end{Rem} 
In what follows, we design the control input $u_i$ for the $i$th, $i=1,\dots,n$ agent such that the output state $x_{i,1}$ can converge towards the reference output $\xi_{i,1}$. To cope with the higher-order dynamics of the agents in (\ref{agent}), the recursive design methodology using the backstepping technique \cite{krstic1995nonlinear} is adopted. We define the tracking errors as follows:
\begin{equation}\label{error}
z_{i,1}=x_{i,1}-\xi_{i,1},
z_{i,\ell}=x_{i,\ell}-\alpha_{i,\ell-1}, \ell=2,\dots,m
\end{equation}
where $\alpha_{i,\ell-1}$ are virtual control signals to be selected. 

\noindent\textit{Step 1:} From (\ref{agent}) and (\ref{observer}), the dynamics of $z_{i,1}$ can be obtained as follows:
\begin{equation}\label{dot_z_1}
\dot{z}_{i,1}=\g_{i,1}(z_{i,2}+\alpha_{i,1})+\theta_i^T\varphi_{i,1}(\bar{x}_{i,1})+\tau_{i,1}(t)-\xi_{i,2}.
\end{equation}
The virtual control $\alpha_{i,1}$ is designed as
\begin{equation}\label{alpha_1}
\alpha_{i,1}=N(k_{i,1})(c_{i,1}z_{i,1}+\hat{\theta}_{i,1}^T\varphi_{i,1}(\bar{x}_{i,1})+\triangle_{i,1}-\xi_{i,2})
\end{equation}
where $N(k_{i,1})=k_{i,1}^2\cos(k_{i,1})$ is the Nussbaum gain, $c_{i,1}$ is a positive constant, $\triangle_{i,1}(t)=\zeta_{i,1}\tanh(z_{i,1}/\varepsilon_{i}(t))$ is a smooth robust term, and $\varepsilon_{i}(t)$ is a positive smooth function satisfying $\int_0^{\infty}\varepsilon_{i}(t)dt\leq \bar{\varepsilon}_{i}$. $\bar{\varepsilon}_{i}$ is a finite positive constant. $\hat{\theta}_{i,1}$ and $\zeta_{i,1}$ represent the estimates of the unknown parameters $\theta_{i}$ and $\bar{\tau}_{i,1}$ with $\bar{\tau}_{i,1}=\tau_{i,1}^*+1$, respectively. The adaptive laws for $\hat{\theta}_{i,1}$, $\zeta_{i,1}$, and $k_{i,1}$ are proposed as 
\begin{equation}\label{estimate}
\begin{array}{c}
\dot{\hat{\theta}}_{i,1}=\varrho_{i,1}\varphi_{i,1}(\bar{x}_{i,1})z_{i,1},
\dot{\zeta}_{i,1}=\mu_{i,1}z_{i,1}\tanh(z_{i,1}/\varepsilon_{i}(t))\\
\dot{k}_{i,1}=(c_{i,1}z_{i,1}+\hat{\theta}_{i,1}^T\varphi_{i,1}(\bar{x}_{i,1})+\triangle_{i,1}-\xi_{i,2})z_{i,1}
\end{array}
\end{equation}
where $\varrho_{i,1}$ and $\mu_{i,1}$ are positive constants. Now, let us consider the following Lyapunov function candidate for the $i$th, $i=1,\dots,n,$ agent as $V_{i,1}=z_{i,1}^2/2+\tilde{\theta}_{i,1}^T\tilde{\theta}_{i,1}/(2\varrho_{i,1})+(\bar{\tau}_{i,1}-\zeta_{i,1})^2/(2\mu_{i,1})$, where $\tilde{\theta}_{i,1}=\theta_{i}-\hat{\theta}_{i,1}$ denotes the parameter estimation error. Its time derivative along (\ref{dot_z_1}) yields
% \begin{equation}\label{dot_V_1}
% \begin{split}
% \dot{V}_{i,1}=&z_{i,1}\dot{z}_{i,1}+\frac{1}{\varrho_{i,1}}\tilde{\theta}_{i,1}^T(-\dot{\hat{\theta}}_{i,1})+\frac{1}{\mu_{i,1}}(\bar{\tau}_{i,1}-\zeta_{i,1})(-\dot{\zeta}_{i,1})\\
% =&\g_{i,1}N(k_{i,1})\dot{k}_{i,1}+(\theta_{i}^T\varphi_{i,1}(\bar{x}_{i,1})+\tau_{i,1}-\xi_{i,2})z_{i,1}
% \\&+\g_{i,1}z_{i,1}z_{i,2}-\frac{\tilde{\theta}_{i,1}^T\dot{\hat{\theta}}_{i,1}}{\varrho_{i,1}}-\frac{1}{\mu_{i,1}}(\bar{\tau}_{i,1}-\zeta_{i,1})\dot{\zeta}_{i,1}.
% \end{split}
% \end{equation}
$\dot{V}_{i,1}
=\g_{i,1}N(k_{i,1})\dot{k}_{i,1}+(\theta_{i}^T\varphi_{i,1}(\bar{x}_{i,1})+\tau_{i,1}-\xi_{i,2})z_{i,1}
+\g_{i,1}z_{i,1}z_{i,2}-\frac{\tilde{\theta}_{i,1}^T\dot{\hat{\theta}}_{i,1}}{\varrho_{i,1}}-\frac{1}{\mu_{i,1}}(\bar{\tau}_{i,1}-\zeta_{i,1})\dot{\zeta}_{i,1}.$
Noting $\dot{k}_{i,1}$ in (\ref{estimate}), we have
\begin{equation}\label{dot_V_2}
\begin{array}{ll}
\dot{V}_{i,1}=&(\g_{i,1}N(k_{i,1})+1)\dot{k}_{i,1}-c_{i,1}z_{i,1}^2+\tilde{\theta}_{i,1}^T\varphi_{i,1}(\bar{x}_{i,1})z_{i,1}\\
&-\triangle_{i,1}z_{i,1}+\tau_{i,1}z_{i,1}+\g_{i,1}z_{i,1}z_{i,2}
\\&
-\frac{\tilde{\theta}_{i,1}^T\dot{\hat{\theta}}_{i,1}}{\varrho_{i,1}}-\frac{1}{\mu_{i,1}}(\bar{\tau}_{i,1}-\zeta_{i,1})\dot{\zeta}_{i,1}.
\end{array}
\end{equation}
Then, substituting adaptive laws (\ref{estimate}) into (\ref{dot_V_2}) results in
\begin{equation}\label{dot_V_21}
\begin{array}{lll}
\dot{V}_{i,1}
% &\leq&(\g_{i,1}N(k_{i,1})+1)\dot{k}_{i,1}-c_{i,1}z_{i,1}^2-z_{i,1}\tanh\left(\frac{z_{i,1}}{\varepsilon_{i}}\right)\\
% &&+\g_{i,1}z_{i,1}z_{i,2}+\kappa_{i,1}\varepsilon_{i}\\
&\leq&(\g_{i,1}N(k_{i,1})+1)\dot{k}_{i,1}-\frac{c_{i,1}z_{i,1}^2}{2}-z_{i,1}\tanh\left(\frac{z_{i,1}}{\varepsilon_{i}}\right)\\
&&+\frac{\g_{i,1}^2z_{i,2}^2}{2c_{i,1}}+\kappa_{i,1}\varepsilon_{i}
\end{array}
\end{equation}
where $|z_{i,1}|\tau^*_{i,1}-z_{i,1}\tau^*_{i,1}\tanh(z_{i,1}/\varepsilon_{i})\leq \kappa_{i,1}\varepsilon_{i}$ with $\kappa_{i,1}=0.2785\tau^*_{i,1}$ \cite{polycarpou1996robust} and Young's inequality $\g_{i,1}z_{i,1}z_{i,2}\leq c_{i,1}z_{i,1}^2/2+\g_{i,1}^2z_{i,2}^2/(2c_{i,1})$ have been used.

\noindent\textit{Step $\ell$ $(2\leq\ell\leq m)$:} A similar procedure is employed recursively for each step. For notational convenience, $z_{i,m+1}=0$ and $u_i=\alpha_{i,m}$ are adopted. The derivative of $z_{i,\ell}$ is computed as
\begin{eqnarray}\label{dot_z_ell}
\begin{array}{lll}
\dot{z}_{i,\ell}
% &=&\g_{i,\ell}(z_{i,\ell+1}+\alpha_{i,\ell})+\theta_i^T\varphi_{i,\ell}(\bar{x}_{i,\ell})+\tau_{i,\ell}-\dot{\alpha}_{i,\ell-1}\\
&=&\g_{i,\ell}(z_{i,\ell+1}+\alpha_{i,\ell})+\theta_{i,\ell}^T\bar{\varphi}_{i,\ell}(\bar{x}_{i,\ell})-\varpi_{i,\ell-1}
\\&&-\sum_{p=1}^{\ell-1}(\partial\alpha_{i,\ell-1}/\partial x_{i,p})\tau_{i,p}+\tau_{i,\ell}
\end{array}
\end{eqnarray}
where $\theta_{i,\ell}=[-\g_{i,1},\dots,-\g_{i,\ell-1},\theta_i^T]^T$ are uncertain parameters, $\bar{\varphi}_{i,\ell}(\bar{x}_{i,\ell})=[(\partial\alpha_{i,\ell-1}/\partial x_{i,1})x_{i,2}, \dots, (\partial\alpha_{i,\ell-1}/\partial x_{i,\ell-1})x_{i,\ell}, [\varphi_{i,\ell}(\bar{x}_{i,\ell})-\sum_{p=1}^{\ell-1}(\partial\alpha_{i,\ell-1}/\partial x_{i,p})\varphi_{i,p}(\bar{x}_{i,p})]^T
 ]^T,$  and $\varpi_{i,\ell-1}=\sum_{p=1}^{\ell-1}[(\partial\alpha_{i,\ell-1}/\partial k_{i,p}) \dot{k}_{i,p}+(\partial\alpha_{i,\ell-1}/\partial\hat{\theta}_{i,p})\dot{\hat{\theta}}_{i,p}+(\partial\alpha_{i,\ell-1}/\partial{\zeta}_{i,p})\dot{\zeta}_{i,p}+(\partial\alpha_{i,\ell-1}/\partial{\varepsilon}_{i}^{(p-1)})\varepsilon_{i}^{(p)}]+\sum_{p=1}^{\ell}(\partial\alpha_{i,\ell-1}/\partial\xi_{i,p})\dot{\xi}_{i,p}.$

% \begin{equation*}
% \begin{array}{lll}
% \varpi_{i,\ell-1}&=&\sum_{p=1}^{\ell-1}[(\partial\alpha_{i,\ell-1}/\partial k_{i,p}) \dot{k}_{i,p}+(\partial\alpha_{i,\ell-1}/\partial\hat{\theta}_{i,p})\dot{\hat{\theta}}_{i,p}
% \\&&+(\partial\alpha_{i,\ell-1}/\partial{\zeta}_{i,p})\dot{\zeta}_{i,p}+(\partial\alpha_{i,\ell-1}/\partial{\varepsilon}_{i}^{(p-1)})\varepsilon_{i}^{(p)}]
% \\&&+\sum_{p=1}^{\ell}(\partial\alpha_{i,\ell-1}/\partial\xi_{i,p})\dot{\xi}_{i,p},
% \end{array}
% \end{equation*}
% \begin{equation*}
% \begin{array}{ll}
% \bar{\varphi}_{i,\ell}(\bar{x}_{i,\ell})=&[(\partial\alpha_{i,\ell-1}/\partial x_{i,1})x_{i,2},\dots,(\partial\alpha_{i,\ell-1}/\partial x_{i,\ell-1})x_{i,\ell},
% \\
% &[\varphi_{i,\ell}(\bar{x}_{i,\ell})-\sum_{p=1}^{\ell-1}(\partial\alpha_{i,\ell-1}/\partial x_{i,p})\varphi_{i,p}(\bar{x}_{i,p})]^T
%  ]^T.
% \end{array}
% \end{equation*}
The virtual control $\alpha_{i,\ell}$ is designed as
\begin{equation}\label{controller}
\alpha_{i,\ell}=N(k_{i,\ell})(c_{i,\ell}z_{i,\ell}+\hat{\theta}_{i,\ell}^T\bar{\varphi}_{i,\ell}(\bar{x}_{i,\ell})+\triangle_{i,\ell}-\varpi_{i,\ell-1})
\end{equation}
where $N(k_{i,\ell})=k_{i,\ell}^2\cos(k_{i,\ell})$ is the Nussbaum gain, $\triangle_{i,\ell}(t)=\zeta_{i,\ell}\eta_{i,\ell}\tanh(\eta_{i,\ell}z_{i,\ell}/\varepsilon_{i}(t))$ with $\eta_{i,\ell}=\sqrt{1+\sum_{p=1}^{\ell-1}(\partial\alpha_{i,\ell-1}/\partial x_{i,p})^2}$, $\hat{\theta}_{i,\ell}$ and $\zeta_{i,\ell}$ are, respectively, the estimates of the unknown parameters $\theta_{i,\ell}$ and $\bar{\tau}_{i,\ell}$ with $\bar{\tau}_{i,\ell}=\sqrt{\ell}\max\{\tau_{i,1}^*,\dots,\tau_{i,\ell}^*\}$. The adaptive laws for $\hat{\theta}_{i,\ell}$, $\zeta_{i,\ell}$, and $k_{i,\ell}$ are selected as 
\begin{equation}\label{estimate_ell}
\begin{array}{c}
\dot{\hat{\theta}}_{i,\ell}=\varrho_{i,\ell}\bar{\varphi}_{i,\ell}(\bar{x}_{i,\ell})z_{i,\ell},
\dot{\zeta}_{i,\ell}=\mu_{i,\ell}\eta_{i,\ell}z_{i,\ell}\tanh(\eta_{i,\ell}z_{i,\ell}/\varepsilon_{i}(t))\\
\dot{k}_{i,\ell}=(c_{i,\ell}z_{i,\ell}+\hat{\theta}_{i,\ell}^T\bar{\varphi}_{i,\ell}(\bar{x}_{i,\ell})+\triangle_{i,\ell}-\varpi_{i,\ell-1})z_{i,\ell}
\end{array}
\end{equation}
where $\varrho_{i,\ell}$ and $\mu_{i,\ell}$ are positive constants. Consider the following Lyapunov function candidate for the $i$th, $i=1,\dots,n,$ agent as $V_{i,\ell}=(1/2)z_{i,\ell}^2+\tilde{\theta}_{i,\ell}^T\tilde{\theta}_{i,\ell}/(2\varrho_{i,\ell})+(\bar{\tau}_{i,\ell}-\zeta_{i,\ell})^2/(2\mu_{i,\ell})$, where $\tilde{\theta}_{i,\ell}=\theta_{i,\ell}-\hat{\theta}_{i,\ell}$ denotes the parameter estimation error. Its time derivative along (\ref{dot_z_ell}) yields
$\dot{V}_{i,\ell}
=\g_{i,\ell}N(k_{i,\ell})\dot{k}_{i,\ell}+(\theta_{i,\ell}^T\bar{\varphi}_{i,\ell}(\bar{x}_{i,\ell})-\varpi_{i,\ell-1})z_{i,,\ell}
+\left(-\sum_{p=1}^{\ell-1}(\partial\alpha_{i,\ell-1}/\partial x_{i,p})\tau_{i,p}+\tau_{i,\ell}\right)z_{i,\ell}
+\g_{i,\ell}z_{i,\ell}z_{i,\ell+1}
-\frac{1}{\varrho_{i,\ell}}\tilde{\theta}_{i,\ell}^T\dot{\hat{\theta}}_{i,\ell}-\frac{1}{\mu_{i,\ell}}(\bar{\tau}_{i,\ell}-\zeta_{i,\ell})\dot{\zeta}_{i,\ell}.$
By $\dot{k}_{i,\ell}$ in (\ref{estimate_ell}), we obtain
\begin{equation}\label{dot_V_ell_2}
\begin{array}{ll}
\dot{V}_{i,\ell}\leq&(\g_{i,\ell}N(k_{i,\ell})+1)\dot{k}_{i,\ell}-c_{i,\ell}z_{i,\ell}^2+\tilde{\theta}_{i,\ell}^T\bar{\varphi}_{i,\ell}(\bar{x}_{i,\ell})z_{i,\ell}\\
&-\triangle_{i,\ell}z_{i,\ell}+\bar{\tau}_{i,\ell}\eta_{i,\ell}|z_{i,\ell}|+\g_{i,\ell}z_{i,\ell}z_{i,\ell+1}
\\&-\frac{1}{\varrho_{i,\ell}}\tilde{\theta}_{i,\ell}^T\dot{\hat{\theta}}_{i,\ell}-\frac{1}{\mu_{i,\ell}}(\bar{\tau}_{i,\ell}-\zeta_{i,\ell})\dot{\zeta}_{i,\ell}
\end{array}
\end{equation}
where the fact that $(\tau_{i,\ell}-\sum_{p=1}^{\ell-1}(\partial\alpha_{i,\ell-1}/\partial x_{i,p})\tau_{i,p})z_{i,\ell}\leq \bar{\tau}_{i,\ell}\eta_{i,\ell}|z_{i,\ell}|$ has been applied. Note that $\bar{\tau}_{i,\ell}\eta_{i,\ell}|z_{i,\ell}|-\bar{\tau}_{i,\ell}\eta_{i,\ell}z_{i,\ell}\tanh(\eta_{i,\ell}z_{i,\ell}/\varepsilon_{i})\leq \kappa_{i,\ell}\varepsilon_{i}$ with $\kappa_{i,\ell}=0.2785\bar{\tau}_{i,\ell}$ \cite{polycarpou1996robust} and that $\g_{i,\ell}z_{i,\ell}z_{i,\ell+1}\leq c_{i,\ell}z_{i,\ell}^2/2+\g_{i,\ell}^2z_{i,\ell+1}^2/(2c_{i,\ell})$. Then, substituting adaptive laws (\ref{estimate_ell}) into (\ref{dot_V_ell_2}) leads to
\begin{equation}\label{dot_V_ell_3}
\dot{V}_{i,\ell}
% &\leq&(\g_{i,\ell}N(k_{i,\ell})+1)\dot{k}_{i,\ell}-c_{i,\ell}z_{i,\ell}^2+\g_{i,\ell}z_{i,\ell}z_{i,\ell+1}+\kappa_{i,\ell}\varepsilon_{i}\\
\leq(\g_{i,\ell}N(k_{i,\ell})+1)\dot{k}_{i,\ell}-\frac{c_{i,\ell}z_{i,\ell}^2}{2}+\frac{\g_{i,\ell}^2z_{i,\ell+1}^2}{2c_{i,\ell}}+\kappa_{i,\ell}\varepsilon_{i}.
\end{equation}

Summarizing the above discussion, we can now state the main result of this brief.
\begin{Thm}\label{Thm1}
Suppose that the directed graph $\mathcal{G}$ contains a spanning tree. Consider a higher-order nonlinear multi-agent system of $n$ agents (\ref{agent}). Under Assumption \ref{asu0}, the proposed distributed control algorithm (\ref{controller}) with the reference output (\ref{observer}) and parameter update laws (\ref{estimate}) and (\ref{estimate_ell}) ensures that (i) all agents can reach asymptotic consensus on the output state, i.e., $\lim_{t\rightarrow\infty}(x_{i,1}(t)-x_{j,1}(t))=0$ for all $1\leq i\neq j\leq n$, and (ii) all signals in the closed-loop system remain bounded.
\end{Thm}

% To prove this theorem, we start by defining $k_i=[k_{i,1},\dots,k_{i,m}]^T$, $\xi_i=[\xi_{i,1},\dots,\xi_{i,m}]^T$, $x_i=\bar{x}_{i,m}$, $\hat{\theta}_i=[\hat{\theta}_{i,1},\dots,\hat{\theta}_{i,m}]^T$, and $\zeta_i=[\zeta_{i,1},\dots,\zeta_{i,m}]^T$ for $i=1,\dots,n$. Then, the closed-loop system can be written in the form of $\dot{\bar{x}}=f(\bar{x},t)$ with the augmented state vector $\bar{x}=[x_1^T,k_1^T,\xi_1^T,\hat{\theta}_1^T,\zeta_{1}^T,\dots,x_n^T,k_n^T,\xi_n^T,\hat{\theta}_n^T,\zeta_n^T]^T\in R^{5n}$. It can be deduced that the map $f(\bar{x},t)$ is locally Lipschitz on $\bar{x}$ and is locally integrable and continuous on $t$ for each fixed $\bar{x}\in R^{5n}$. Thus, according to \cite[Th. 54]{sontag2013mathematical}, a unique solution exists within some time interval $[0,t_f)$.
\textit{Proof} In view of $z_{i,m+1}=0$, we can obtain from (\ref{dot_V_ell_3}) that $\dot{V}_{i,m}
\leq(\g_{i,m}N(k_{i,m})+1)\dot{k}_{i,m}-c_{i,m}z_{i,m}^2+\kappa_{i,m}\varepsilon_{i}.$
Integrating this inequality between $0$ and $t$ leads to
\begin{equation}\label{int_V_m_a}
\begin{array}{lll}
V_{i,m}(t)
&\leq&V_{i,m}(0)+\int_0^t(\g_{i,m}N(k_{i,m}(\sigma))+1)\dot{k}_{i,m}(\sigma)d\sigma\\
&&-\int_0^tc_{i,m}z_{i,m}^2(\sigma)d\sigma+\int_0^t\kappa_{i,m}\varepsilon_i(\sigma)d\sigma\\
&\leq&\bar{V}_{i,m}+\int_0^t(\g_{i,m}N(k_{i,m}(\sigma))+1)\dot{k}_{i,m}(\sigma)d\sigma
\end{array}
\end{equation}
where $\bar{V}_{i,m}=V_{i,m}(0)+\kappa_{i,m}\bar{\varepsilon}_i$. Applying \cite[Lem. 1]{xudong1998adaptive} to (\ref{int_V_m_a}), we conclude that $V_{i,m},k_{i,m}\in\mathcal{L}_\infty[0,t_f)$; furthermore, $z_{i,m}\in\mathcal{L}_2[0,t_f)$ and $z_{i,m},\tilde{\theta}_{i,m},(\bar{\tau}_{i,m}-\zeta_{i,m})\in\mathcal{L}_\infty[0,t_f)$. The boundedness of $\hat{\theta}_{i,m}(t)$ and $\zeta_{i,m}(t)$ on $[0,t_f)$ then follows from the fact that $\theta_{i,m}$ and $\bar{\tau}_{i,m}$ are constants. Employing \cite[Lem. 1]{xudong1998adaptive} recursively $(m-2)$ times, it can be shown from the aforementioned design procedures that $V_{i,\ell},k_{i,\ell},z_{i,\ell},\hat{\theta}_{i,\ell},\zeta_{i,\ell}\in\mathcal{L}_\infty[0,t_f)$ and $z_{i,\ell}\in\mathcal{L}_2[0,t_f)$ for all $\ell=2,\dots,m-1$. Since $z_{i,2}\in\mathcal{L}_2[0,t_f)$, there exists a positive constant $\bar{b}_i$ such that $\int_0^t \g_{i,1}^2z_{i,2}^2(\sigma)/(2c_{i,1})d\sigma\leq \bar{b}_i$ for all $t\in[0,t_f)$. Integrating (\ref{dot_V_21}) between $0$ and $t$, $\forall t\in[0,t_f)$ gives $V_{i,1}(t)\leq V_{i,1}(0)+\int_0^t(\g_{i,1}N(k_{i,1}(\sigma))+1)\dot{k}_{i,1}(\sigma)d\sigma
-\int_0^t\frac{c_{i,1}z_{i,1}^2(\sigma)}{2}d\sigma+\int_0^t\kappa_{i,1}\varepsilon_i(\sigma)d\sigma+\bar{b}_i
-\int_0^tz_{i,1}(\sigma)\tanh\left(\frac{z_{i,1}(\sigma)}{\varepsilon_{i}(\sigma)}\right)d\sigma
\leq\bar{V}_{i,1}+\int_0^t(\g_{i,1}N(k_{i,1}(\sigma))+1)\dot{k}_{i,1}(\sigma)d\sigma,$ where $\bar{V}_{i,1}=V_{i,1}(0)+\kappa_{i,1}\bar{\varepsilon}_i+\bar{b}_i$ is a finite positive constant. Thus, we can conclude from \cite[Lem. 1]{xudong1998adaptive} that $V_{i,1},k_{i,1},z_{i,1},\hat{\theta}_{i,1},\zeta_{i,1}\in\mathcal{L}_\infty[0,t_f)$, which implies that $z_{i,1}\in\mathcal{L}_2[0,t_f)$ and $\int_0^tz_{i,1}(\sigma)\tanh(z_{i,1}(\sigma)/\varepsilon_i(\sigma))d\sigma$ is bounded on $[0,t_f)$. Because $0\leq \int_0^t|z_{i,1}(\sigma)|d\sigma\leq\int_0^tz_{i,1}(\sigma)\tanh(z_{i,1}(\sigma)/\varepsilon_i(\sigma))d\sigma
+\int_0^t 0.2785\varepsilon_i(\sigma)d\sigma$ and $\varepsilon_i\in\mathcal{L}_1[0,\infty)$, we can obtain $z_{i,1}\in\mathcal{L}_1[0,t_f)$.

Noticing (\ref{observer}), the dynamics of the reference output can be rewritten as $\dot{\xi}_i=A_i\gamma_i\sum_{j=1}^na_{ij}x_{j,1}+B_i\xi_i,$ where $i\in\mathcal{V}_1$, $\xi_i=[\xi_{i,1},\dots,\xi_{i,m}]^T$, $A_i=[0_{m-1}^T,1]^T$,  and 
$B_i=\left[\begin{array}{cc}
0_{m-1}&I_{m-1}\\
-\lambda_{i,1}&-\bar{\lambda}_i
\end{array}\right]$ with $\bar{\lambda}_i=[\lambda_{i,2},\dots,\lambda_{i,m}].$ Let the eigenvalues of $B_i$ be denoted, without a particular order, by $\delta_{i,\ell}<0$ for $\ell=1,\dots,m$. For analysis purposes, we introduce a nonsingular transformation matrix $J_i$ for the $i$th agent as \cite{abdessameud2018distributed}
\begin{equation*}
J_i=\left[
\begin{array}{ccccc}
1&0&\dots&0&0\\
1&j_{i,22}&\dots&0&0\\
% 1&j_{i,32}&j_{i,33}&\dots&0&0\\
\vdots&\vdots&\ddots&\vdots&\vdots\\
1&j_{i,(m-1)2}&\dots&j_{i,(m-1)(m-1)}&0\\
1&j_{i,m2}&\dots&j_{i,m(m-1)}&j_{i,mm}\\
\end{array}
\right].
\end{equation*}
where $j_{i,\ell p}=j_{i,(\ell-1)p}-j_{i,(\ell-1)(p-1)}/\delta_{i,\ell-1}$ for $\ell=2,\dots,m$, $2\leq p\leq\ell$. Then, it can be verified that $J_iB_i=\Lambda_iJ_i$ and $j_{i,mm}=-\delta_{i,m}/\lambda_{i,1}$ hold, where $\Lambda_i=\left[
\begin{array}{ccccc}
\delta_{i,1} &-\delta_{i,1}&0&\dots&0\\
0 &\delta_{i,2}&-\delta_{i,2}&\dots&0\\
\vdots &\vdots&\vdots&\ddots&\vdots\\
0 &0&0&\dots&\delta_{i,m}
\end{array}\right].$ Taking the state transformation $q_i=[q_{i,1},\dots,q_{i,m}]^T=J_i\xi_i$ for $i\in\mathcal{V}_1$, and noting (\ref{error}), we have 
\begin{eqnarray}\label{observer_dy}
\begin{array}{lll}
\dot{q}_{i,\ell}&=&\delta_{i,1}q_{i,\ell}-\delta_{i,1}q_{i,\ell+1},\ell=1,\dots,m-1\\
\dot{q}_{i,m}&=&\delta_{i,m}q_{i,m}-(\delta_{i,m}/d_i)\sum_{j=1}^na_{ij}(q_{j,1}+z_{j,1}).
\end{array}
\end{eqnarray}

Since the directed graph $\mathcal{G}$ contains a spanning tree, there is at most one agent with no neighbors. We consider two cases: (\textit{C1}) each agent can obtain information from at least one other agent, i.e., $\mathcal{V}_1=\mathcal{V}$ and (\textit{C2}) there exists one agent that cannot receive any information from any other agent.

\textit{C1:} Define the column vectors $\bar{q}_{\ell}=[q_{1,\ell},\dots,q_{n,\ell}]^T$, $z=[0_{(m-1)n}^T,\bar{z}_{1,1},\dots,\bar{z}_{n,1}]^T$, and $q=[\bar{q}_1^T,\dots,\bar{q}_m^T]^T$, where $\bar{z}_{i,1}=-(\delta_{i,m}/d_i)\sum_{j=1}^na_{ij}z_{j,1}$ for $i=1,\dots,n$, $\ell=1,\dots,m$. It follows from (\ref{observer_dy}) that
\begin{equation}\label{d_q}
\dot{q}(t)=-{\bar{L}}q(t)+z(t)
\end{equation}
where
\begin{equation}\label{bar_L}
{\bar{L}}=
\left[
\begin{array}{cccccc}
-\delta_1&\delta_1&0 &\dots&0&0\\
% 0&-\delta_2&\delta_2 &\dots&0&0\\
\vdots&\vdots&\vdots &\ddots&\vdots&\vdots\\
0&0&0 &\dots&-\delta_{m-1}&\delta_{m-1}\\
\delta_m{\bar{A}}_n&0&0 &\dots&0&-\delta_m
\end{array}\right]
\end{equation}
$\delta_\ell=\diag\{\delta_{1,\ell},\dots,\delta_{n,\ell}\}$ and ${\bar{A}}_n={D}_n^{-1}{A}_n$. Apparently, the matrix ${\bar{L}}$ is in the form of a Laplacian matrix, as ${\bar{L}}1_{nm}=0_{nm}$, and all the off-diagonal entries of ${\bar{L}}$ are non-positive. Consequently, the dynamic system (\ref{d_q}) can be treated as a system consisting of $nm$ agents that are interconnected according to the augmented directed graph $\mathcal{\bar{G}}=(\mathcal{\bar{V}},\mathcal{\bar{E}})$, in which $\mathcal{\bar{V}}=\{1,\dots,nm\}$, ${\bar{L}}$ is the associated Laplacian matrix, and the edge set $\mathcal{\bar{E}}$ can be decided from (\ref{bar_L}). Note that $\rank(\delta_{\ell})=n$ and $\rank(\delta_m(I_n-{\bar{A}}_n))=\rank({L}_n)=n-1$. Applying Lemma \ref{Lemma1} to (\ref{bar_L}), we obtain that $\rank({\bar{L}})=nm-1$. Hence, it follows from \cite{renbook2008distributed} that ${\bar{L}}$ has a single zero eigenvalue and all other eigenvalues have positive real parts, which means that there exists a finite constant $\bar{m}$ such that $\|e^{-{\bar{L}}t}\|\leq \bar{m}$ for all $t\geq 0$ \cite[p. 138]{chen1998linear}. Integrating both sides of (\ref{d_q}) with respect to $t$ gives the solution $q(t)=e^{-{\bar{L}}t}q(0)+\int_0^te^{-{\bar{L}}(t-\sigma)}z(\sigma)d\sigma$. By utilizing $\|e^{-{\bar{L}}t}\|\leq \bar{m}$ and $z_{i,1}\in\mathcal{L}_1[0,t_f)$, we have $q\in\mathcal{L}_\infty[0,t_f)$. Then, we can verify that $\xi_i\in\mathcal{L}_\infty[0,t_f)$ since $J_i$ is nonsingular. Noting $z_{i,1}=x_{i,1}-\xi_{i,1}$ and $z_{i,1}\in\mathcal{L}_\infty[0,t_f)$, we can conclude that $x_{i,1}\in\mathcal{L}_\infty[0,t_f)$. Since $\varphi_{i,1}$ is a smooth nonlinear function on $x_{i,1}$, we obtain from (\ref{alpha_1}) and (\ref{estimate}) that $\alpha_{i,1},\dot{\theta}_{i,1},\dot{\zeta}_{i,1},\dot{k}_{i,1}\in\mathcal{L}_\infty[0,t_f)$. Using the mimicking argument, $x_{i,\ell},\alpha_{i,\ell},\dot{\theta}_{i,\ell},\dot{\zeta}_{i,\ell},\dot{k}_{i,\ell}\in\mathcal{L}_\infty[0,t_f)$ for $\ell=2,\dots,m$. Therefore, no finite-time escape phenomenon may occur, and the solution can be extended to $t_f=\infty$. From (\ref{dot_z_1}) and (\ref{dot_z_ell}), we have $\dot{z}_{i,p}\in\mathcal{L}_\infty[0,\infty)$ for $i=1,\dots,n,p=1,\dots,m$. Combining this with $z_{i,p}\in\mathcal{L}_\infty[0,\infty)\cap\mathcal{L}_2[0,\infty)$, we conclude from Barbalat's lemma that $\lim_{t\rightarrow\infty}z_{i,p}(t)=0$.

In the sequel, we show that asymptotic consensus can be achieved. To this end, we introduce the relative error vectors $\tilde{q}=[\underline{q}_1-\underline{q}_2,\dots,\underline{q}_{nm-1}-\underline{q}_{nm}]^T\in R^{nm-1}$ and $\tilde{z}=[\underline{z}_1-\underline{z}_2,\dots,\underline{z}_{nm-1}-\underline{z}_{nm}]^T\in R^{nm-1}$, where $\underline{q}_p$ and $\underline{z}_p$ are, respectively, the $p$th element of $q$ and $z$ for $p=1,\dots,nm$. The dynamics of $\tilde{q}$ can be obtained from (\ref{d_q}) as $\dot{\tilde{q}}=-\Omega\tilde{q}+\tilde{z}$, where $\Omega\in R^{(nm-1)\times(nm-1)}$ is time-invariant. It follows from $\rank({\bar{L}})=nm-1$ and \cite{renbook2008distributed} that $\mathcal{\bar{G}}$ contains a spanning tree. By \cite[Th. 2.14]{renbook2008distributed}, we can show that the system $\dot{\tilde{q}}=-\Omega\tilde{q}$ is asymptotically stable. Note from \cite[Th. 4.14]{panos} that if a linear time-invariant system is asymptotically stable, then it is also exponentially stable. Combining this with the fact that $z_{i,1}\in\mathcal{L}_1[0,\infty)$, we obtain $\lim_{t\rightarrow\infty}\tilde{q}(t)=0_{nm-1}$. Recalling that $\xi_{i,1}=q_{i,1}$, we have $\lim_{t\rightarrow\infty}(\xi_{i,1}(t)-\xi_{j,1}(t))=0$, which together with $\lim_{t\rightarrow\infty}(x_{i,1}(t)-\xi_{i,1}(t))=0$ yields that $\lim_{t\rightarrow\infty}(x_{i,1}(t)-x_{j,1}(t))=0$ for all $i,j=1,\dots,n$.

\textit{C2:} In this case, we assume, without loss of generality, that the node indexed by $1$ is the agent having no neighbors. We use $\underline{\mathcal{G}}$ with the node set $\underline{\mathcal{V}}=\{2,\dots,n\}$ and the edge set $\underline{\mathcal{E}}\subseteq\underline{\mathcal{V}}\times \underline{\mathcal{V}}$ to describe the communication topology among the agents $2$ to $n$. Let ${A}_{n-1}$, ${D}_{n-1}$, and ${L}_{n-1}$ denote, respectively, the weighted adjacency matrix, the in-degree matrix, and the Laplacian matrix associated with $\mathcal{G}_{n-1}$. Then, the Laplacian matrix ${L}_n$ associated with $\mathcal{G}$ can be partitioned as
${L}_n=\left[
\begin{array}{cc}
0&0_{n-1}^T\\
h&{L}_{n-1}
\end{array}
\right],$ where $h=[a_{21},\dots,a_{n1}]^T\in R^{n-1}$. Since $\mathcal{G}$ contains a spanning tree, it follows from \cite{renbook2008distributed} that $\rank({L})=n-1$, which indicates that $\rank({L}_{n-1})=n-1$.

Define the column vectors $\bar{q}_{1}=[\xi_{1,1},q_{2,1},\dots,q_{n,1}]^T$, $\bar{q}_{\ell}=[q_{2,\ell},\dots,q_{n,\ell}]^T$, $z=[0_{(n-1)(m-1)+1}^T,\bar{z}_{2,1},\dots,\bar{z}_{n,1}]^T$, and $q=[\bar{q}_1^T,\dots,\bar{q}_m^T]^T$, where $i=2,\dots,n,$ $\ell=2,\dots,m,$ and $\bar{z}_{i,1}=-(\delta_{i,m}/d_i)\sum_{j=1}^na_{ij}z_{j,1}$. Noting (\ref{observer_dy}) and the fact that $\xi_{1,1}$ is a constant, we have
\begin{equation}\label{d_q_new}
\dot{q}(t)=-{\bar{L}}q(t)+z(t)
\end{equation}
where
\begin{equation}\label{L_27}
{\bar{L}}=\left[
\begin{array}{cc}
0&0_{(n-1)m}^T\\
\hbar&{L}_{(n-1)m}
\end{array}
\right]
\end{equation}
with $\hbar=[0_{(n-1)(m-1)}^T,h^T]^T\in R^{(n-1)m}$,
\begin{equation*}
{L}_{(n-1)m}=
\left[
\begin{array}{cccccc}
-\delta_1&\delta_1&0 &\dots&0&0\\
% 0&-\delta_2&\delta_2 &\dots&0&0\\
\vdots&\vdots&\vdots &\ddots&\vdots&\vdots\\
0&0&0 &\dots&-\delta_{m-1}&\delta_{m-1}\\
\delta_m{\bar{A}}_{n-1}&0&0 &\dots&0&-\delta_m
\end{array}\right]
\end{equation*}
$\delta_\ell=\diag\{\delta_{2,\ell},\dots,\delta_{n,\ell}\}$ for $\ell=1,\dots,m$, and ${\bar{A}}_{n-1}={D}_{n-1}^{-1}{A}_{n-1}$. Notice that the matrix ${\bar{L}}$ takes the form of a Laplacian matrix, as ${\bar{L}}1_{(n-1)m+1}=0_{(n-1)m+1}$ and all the off-diagonal entries of ${\bar{L}}$ are non-positive. The system (\ref{d_q_new}) can be regarded as a system consisting of $(n-1)m+1$ agents that are interconnected according to the augmented directed graph $\mathcal{\bar{G}}=(\mathcal{\bar{V}},\mathcal{\bar{E}})$, where $\mathcal{\bar{V}}=\{1,\dots,(n-1)m+1\}$, ${\bar{L}}$ is the associated Laplacian matrix, and the edge set $\mathcal{\bar{E}}$ can be determined from (\ref{L_27}). Since $\rank(\delta_\ell)=n-1$ and $\rank(\delta_m(I_{n-1}-{\bar{A}}_{n-1}))=\rank({L}_{n-1})=n-1$, it follows from Lemma \ref{Lemma1} that $\rank({L}_{(n-1)m})=(n-1)m$. By (\ref{L_27}), we have $\rank({\bar{L}})=(n-1)m$. Now proceeding in a manner similar to the proof of \textit{C1}, we can conclude that all signals in the closed-loop system are bounded and that $\lim_{t\rightarrow\infty}(\xi_{i,1}(t)-\xi_{j,1}(t))=0$, $\lim_{t\rightarrow\infty}(x_{i,1}(t)-\xi_{i,1}(t))=0$, and $\lim_{t\rightarrow\infty}(x_{i,1}(t)-x_{j,1}(t))=0$ for all $i,j=1,\dots,n$. Furthermore, since $\xi_{i,1}$ is a constant for $i\in\mathcal{V}_2$ in this case, we can also conclude that $\lim_{t\rightarrow\infty}x_{j,1}(t)=\gamma_i$ for all $j=1,\dots,n$, which completes the proof.

\section{Simulation study}
\begin{figure}
\begin{center}
\includegraphics[height=2cm]{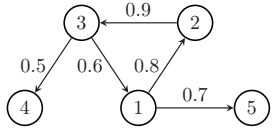}    % The printed column    % width is 8.4 cm.
\caption{Directed communication topology.}
\label{fig_topo}   % width is 8.4 cm.
\end{center}
\end{figure}
\textbf{Example 1:} An application example with five marine vessels is considered to clarify and verify the theoretical findings of our work. The Norrbin model for the $i$th, $i = 1,\dots,5$, vessel (agent) can be described by \cite{fossen2011handbook}
\begin{equation*}
T_i\ddot{\vartheta}_i+\dot{\vartheta}_i+W_i\dot{\vartheta}_i^3=M_i\psi_i+\rho_i(t)
\end{equation*}
where $\vartheta_i$ denotes the actual course of the $i$th vessel,  $\psi_i$ is the rudder angle, $M_i$ is the gain constant, $T_i$ is the time constant, $W_i$ is the Norrbin coefficient, and $\rho_i(t)$ denotes the time-varying disturbance term. The control objective is to design the distributed rudder angle $\psi_i$ for the $i$th vessel by utilizing only its states $\vartheta_i$, $\dot{\vartheta}_i$ and the course $\vartheta_j$ $(j\in\mathcal{N}_i)$ of its neighbors so that the course of all vessels can achieve asymptotic consensus. Note that since only the course (output information) of the neighbors is available, the results in \cite{chen2014adaptive,pengj2018cooperative,chen2017adaptive,wang2019distributed} cannot be applied to realize this goal. We define $x_{i,1}=\vartheta_i$, $x_{i,2}=\dot{\vartheta}_i$, and $u_i=\psi_i$. Then, the dynamics of the vessel can be equivalently written as $\dot{x}_{i,1}=x_{i,2},$ $\dot{x}_{i,2}=\g_{i}u_i+\theta_i^T\varphi_i(\bar{x}_{i,2})+\tau_i(t),$ where $\bar{x}_{i,2}=[x_{i,1},x_{i,2}]^T$, $\g_i=M_i/T_i$, $\varphi_i(\bar{x}_{i,2})=[x_{i,2},x_{i,2}^3]^T$, $\theta_i=[-1/T_i, -W_i/T_i]^T$, and $\tau_i(t)=\rho_i(t)/T_i$. The simulation parameters are set as follows: $W_i=0.4-0.05i$, $M_i=20+0.5i$, $T_i=(-1)^i+0.02i$, and $\rho_i=0.2i\cos(t)$. The initial postures of vessels are $\bar{x}_{1,2}(0)=[\pi/2,0]^T$, $\bar{x}_{2,2}(0)=[-\pi/4,0]^T$, $\bar{x}_{3,2}(0)=[-\pi/3,0]^T$, $\bar{x}_{4,2}(0)=[\pi/5,0]^T$, and $\bar{x}_{5,2}(0)=[-\pi/2,0]^T$. We consider two cases: (\textit{S1}) each vessel can obtain information from at least one other vessel, and (\textit{S2}) there exists one vessel that cannot receive any information from any other vessel.

\begin{figure}
\begin{center}
\includegraphics[height=5.5cm]{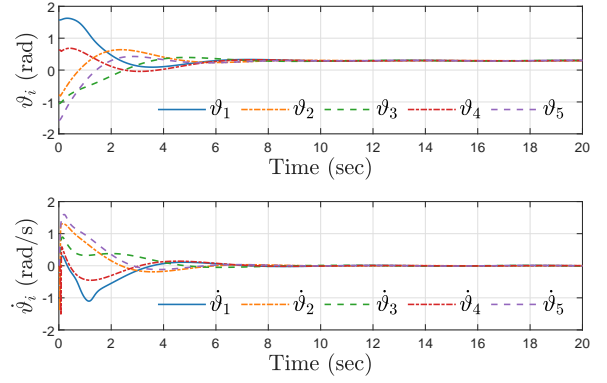}    % The printed column    % width is 8.4 cm.
\caption{Trajectories of the course $\vartheta_i$ and velocities $\dot{\vartheta}_i$ for $1\leq i\leq 5$ under the proposed algorithm. All vessels reach asymptotic consensus on the course, i.e., $\lim_{t\rightarrow\infty}(\vartheta_i(t)-\vartheta_j(t))=0$, $\forall 1\leq i\neq j\leq 5$.}
\label{fig_x1}   % width is 8.4 cm.
\end{center}
\end{figure}

\textit{S1:} The communication topology among the vessels is shown in Fig. \ref{fig_topo}. Since $\g_{i,1}=1$ and $\varphi_{i,1}(x_{i,1})=0$ for $i=1,\dots,5$, we choose the virtual control $\alpha_{i,1}$ as $\alpha_{i}=-c_{i,1}z_{i,1}-\triangle_{i,1}+\xi_{i,2}$. The control input $u_i$ is designed as $u_i=N(k_{i})(c_{i,2}z_{i,2}+\hat{\theta}_{i}^T\varphi_{i}(\bar{x}_{i,2})+\triangle_{i,2}-\dot{\alpha}_{i})$. The initials for $\hat{\theta}_i$, $\zeta_{i,1}$, $\zeta_{i,2}$, and $k_i$ are all set as zero. $\xi_{i,1}(0)$ is chosen randomly in the range of $[-\pi/2,\pi/2]$ and $\xi_{i,2}(0)=0$. The design parameters for the proposed algorithm are chosen as $c_{i,1}=c_{i,2}=4$, $\mu_{i,1}=\mu_{i,2}=0.2$, $\lambda_{i,1}=1$, $\lambda_{i,2}=2$, $\varrho_{i}=2$, and $\varepsilon_{i}=e^{-0.05t}$.  Fig. \ref{fig_x1} depicts the simulation results. Note that all vessels achieve consensus asymptotically with the developed control algorithm, and the velocities $x_{i,2}$ are bounded. 

\begin{figure}
\begin{center}
\includegraphics[height=2cm]{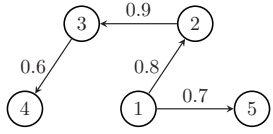}    % The printed column    % width is 8.4 cm.
\caption{Directed communication topology.}
\label{fig_topo_2}   % width is 8.4 cm.
\end{center}
\end{figure}

\begin{figure}
\begin{center}
\includegraphics[height=5.5cm]{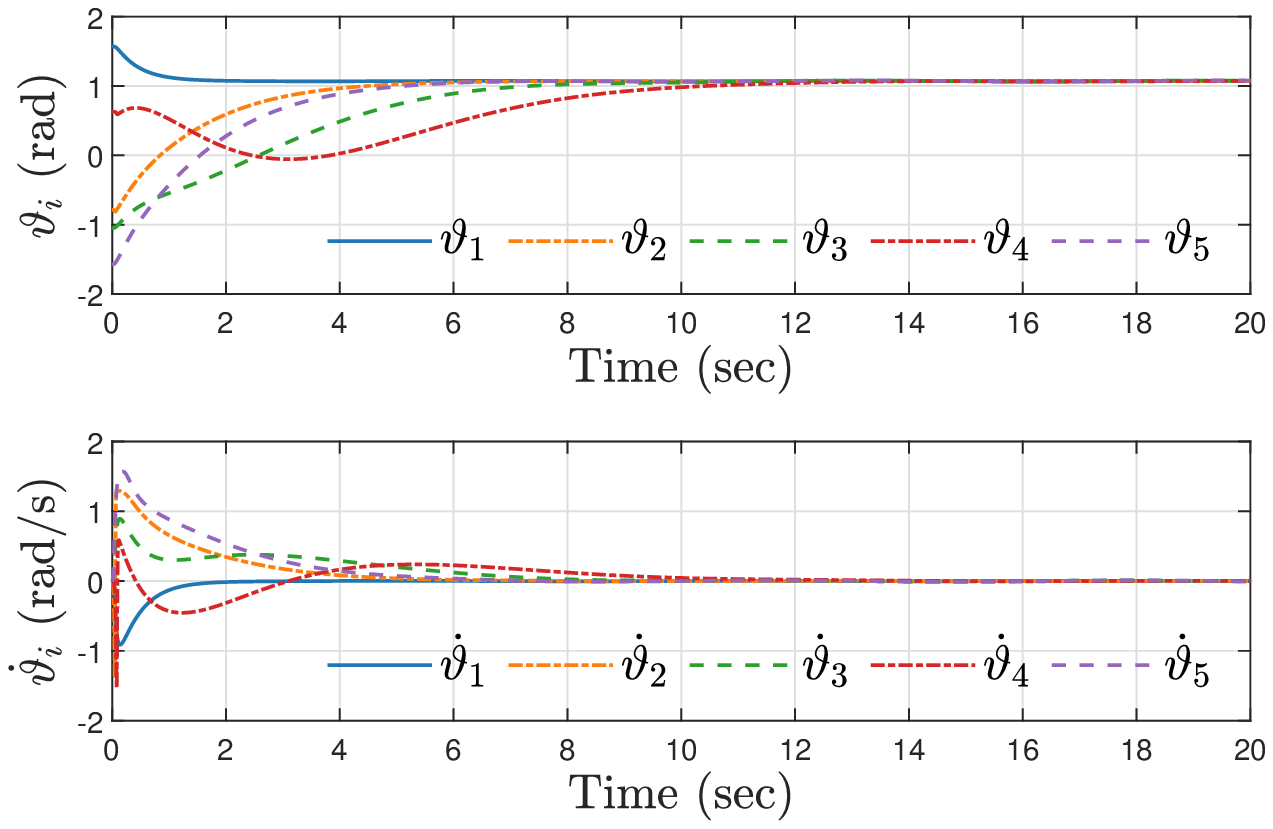}    % The printed column    % width is 8.4 cm.
\caption{Trajectories of the course $\vartheta_i$ and velocities $\dot{\vartheta}_i$ for $1\leq i\leq 5$ under the proposed algorithm. All vessels reach asymptotic consensus on the course, i.e., $\lim_{t\rightarrow\infty}\vartheta_i(t)=\gamma_1$ for $1\leq i\leq 5$.}
\label{fig_case2}   % width is 8.4 cm.
\end{center}
\end{figure}

\textit{S2:} In this case, the communication topology among the vessels is shown in Fig. \ref{fig_topo_2}.  Since the vessel indexed by 1 cannot receive any information from any other vessel, its reference output $\xi_{1,1}$ is designed as $\xi_{1,1}=\gamma_1$, where $\gamma_1$ is a randomly selected constant in the range of $[-\pi/2,\pi/2]$. Control input design, the rest of the reference output and parameter settings are the same as in \textit{S1}. The profiles of each vessel's angle and velocity are exhibited in Fig. \ref{fig_case2}.

\textbf{Example 2:} Consider that there is a group of five second-order nonlinear agents with the dynamics
\begin{equation*}
\begin{array}{cll}
\dot{x}_{i,1}&=&\g_{i,1}x_{i,2}+\theta_{i}^T\varphi_{i,1}({x}_{i,1})+\tau_{i,1}(t) \\
\dot{x}_{i,2}&=&\g_{i,2}u_i+\theta_{i}^T\varphi_{i,2}(\bar{x}_{i,2})+\tau_{i,2}(t)
\end{array}
\end{equation*}
where $i=1,\dots,5$, $\varphi_{i,1}({x}_{i,1})=\cos(x_{i,1})$, $\varphi_{i,2}(\bar{x}_{i,2})=x_{i,1}\sin(x_{i,2})$, $\g_{i,1}=1.1-0.1i$, $\g_{i,2}=(-1)^i(0.5+0.1i)$, $\theta_i=1-0.1i$, $\tau_{i,1}(t)=(0.6-0.1i)\sin(t)$, and $\tau_{i,2}(t)=0.1i\cos(t)$. The communication topology among the agents is shown in Fig. \ref{fig_topo}. The control goal is to design consensus algorithms for agents based only on their states and the output information of their neighbors so that all agents can reach asymptotic consensus on the output state, i.e., $\lim_{t\rightarrow\infty}(x_{i,1}(t)-x_{j,1}(t))=0$ for all $1\leq i\neq j\leq 5$. 

\begin{figure}
\begin{center}
\includegraphics[height=5.5cm]{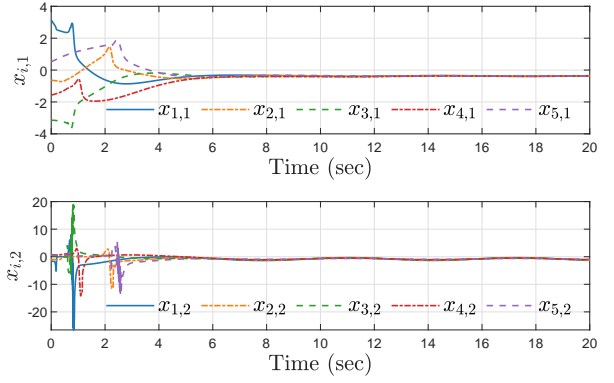}    % The printed column    % width is 8.4 cm.
\caption{Trajectories of $x_{i,1}$ and $x_{i,2}$ for $1\leq i\leq 5$ under the proposed algorithm. All agents reach asymptotic consensus on the output, i.e., $\lim_{t\rightarrow\infty}(x_{i,1}(t)-x_{j,1}(t))=0$, $\forall 1\leq i\neq j\leq 5$.}
\label{fig_case3}   % width is 8.4 cm.
\end{center}
\end{figure}

In the design procedure, the error variables for each agent are defined as
$z_{i,1}=x_{i,1}-\xi_{i,1}$ and $z_{i,2}=x_{i,2}-\alpha_{i,1}$, where $\xi_{i,1}$ is selected by (\ref{observer}). By (\ref{alpha_1}), we design the virtual control $\alpha_{i,1}$ as $\alpha_{i,1}=N(k_{i,1})(c_{i,1}z_{i,1}+\hat{\theta}_{i,1}^T\varphi_{i,1}(x_{i,1})+\triangle_{i,1}-\xi_{i,2})$. According to (\ref{controller}), the control input $u_i$ is designed as $u_i=N(k_{i,2})(c_{i,2}z_{i,2}+\hat{\theta}_{i,2}^T\bar{\varphi}_{i,2}(\bar{x}_{i,2})+\triangle_{i,2}-\varpi_{i,1})$, where $\varpi_{i,1}=(\partial\alpha_{i,1}/\partial k_{i,1}) \dot{k}_{i,1}+(\partial\alpha_{i,1}/\partial\hat{\theta}_{i,1})\dot{\hat{\theta}}_{i,1}+(\partial\alpha_{i,1}/\partial{\zeta}_{i,1})\dot{\zeta}_{i,1}+(\partial\alpha_{i,1}/\partial{\varepsilon}_{i})\dot{\varepsilon}_{i}+\sum_{p=1}^{2}(\partial\alpha_{i,1}/\partial\xi_{i,p})\dot{\xi}_{i,p}$ and $\bar{\varphi}_{i,2}(\bar{x}_{i,2})=[(\partial\alpha_{i,1}/\partial x_{i,1})x_{i,2}, \varphi_{i,2}(\bar{x}_{i,2})-(\partial\alpha_{i,1}/\partial x_{i,1})\varphi_{i,1}(x_{i,1})]^T.$ Parameter update laws are designed as
\begin{equation*}
\begin{array}{c}
\dot{\hat{\theta}}_{i,1}=\varrho_{i,1}\varphi_{i,1}(x_{i,1})z_{i,1},
\dot{\zeta}_{i,1}=\mu_{i,1}z_{i,1}\tanh(z_{i,1}/\varepsilon_{i}(t))\\
\dot{k}_{i,1}=(c_{i,1}z_{i,1}+\hat{\theta}_{i,1}^T\varphi_{i,1}(x_{i,1})+\triangle_{i,1}-\xi_{i,2})z_{i,1}\\
\dot{\hat{\theta}}_{i,2}=\varrho_{i,2}\bar{\varphi}_{i,2}(\bar{x}_{i,2})z_{i,2},
\dot{\zeta}_{i,2}=\mu_{i,2}\eta_{i,2}z_{i,2}\tanh(\eta_{i,2}z_{i,2}/\varepsilon_{i}(t))\\
\dot{k}_{i,2}=(c_{i,2}z_{i,2}+\hat{\theta}_{i,2}^T\bar{\varphi}_{i,2}(\bar{x}_{i,2})+\triangle_{i,2}-\varpi_{i,1})z_{i,2}
\end{array}
\end{equation*}
where $\eta_{i,2}=\sqrt{1+(\partial\alpha_{i,1}/\partial x_{i,1})^2}$. The initial conditions are given by $\bar{x}_{1,2}(0)=[\pi,-\pi/2]^T$, $\bar{x}_{2,2}(0)=[-\pi/5,-\pi/3]^T$, $\bar{x}_{3,2}(0)=[-\pi,\pi/4]^T$, $\bar{x}_{4,2}(0)=[-\pi/2,\pi/5]^T$, and $\bar{x}_{5,2}(0)=[\pi/6,\pi/6]^T$.
All the estimate initials are set as zero, except that $\xi_i(0)=\bar{x}_{i,2}(0)$. The controller coefficients are chosen as $c_{i,1}=c_{i,2}=5$, $\mu_{i,1}=\mu_{i,2}=0.5$, $\lambda_{i,1}=1$, $\lambda_{i,2}=2$, $\varrho_{i,1}=\varrho_{i,1}=1$, and $\varepsilon_{i}=e^{-0.05t}$.  Fig. \ref{fig_case3} depicts the simulation results. Through the simulation results, the proposed distributed adaptive algorithm can fulfill the consensus assignment under a directed graph condition and has satisfactory control performance, despite the presence of mismatched uncertainties and multiple completely unknown high-frequency gains.

\section{Conclusion}
In this brief, a distributed control algorithm for nonlinear multi-agent systems with mismatched uncertainties and unknown high-frequency gains under a directed graph was proposed. One salient feature of the presented algorithm is that it requires minimal information from neighboring agents, namely, their output measurements, such that asymptotic consensus among agents can be reached with lower communication costs. This overcomes a typical problem of the existing consensus schemes on higher-order nonlinear agents, in which both the states of the neighbors and their preliminary dynamics knowledge are needed for each agent to design the control input. Simulation results on marine vessels validated the theoretical finding. Following the metaheuristic algorithm presented in \cite{khan2020tracking,hamzatracking,khan2019obstacle}, future work will be devoted to handling the consensus problem of redundant robotic manipulators.

\bibliographystyle{IEEEtran}

\bibliography{IEEEabrv}

\end{document}